\documentclass[prl,aps,twocolumn,showpacs,superscriptaddress,floatfix]{revtex4}
\usepackage{graphicx}
\usepackage{amsmath}

\begin{document}

\title{Conductance of fully equilibrated quantum wires}

\author{J\'er\^ome Rech} 

\affiliation{Materials Science Division, Argonne National Laboratory,
  Argonne, Illinois 60439, USA}

\affiliation{Physics Department, Arnold Sommerfeld Center for Theoretical
  Physics, and Center for NanoScience, Ludwig-Maximilians-Universit\"at,
  Theresienstrasse 37, 80333 Munich, Germany}

\author{Tobias Micklitz} 

\affiliation{Materials Science Division, Argonne National Laboratory,
  Argonne, Illinois 60439, USA}

\author{K. A. Matveev} 

\affiliation{Materials Science Division, Argonne National Laboratory,
  Argonne, Illinois 60439, USA}

\date{\today} 

\pacs{71.10.Pm}

\begin{abstract}

  We study the conductance of a quantum wire in the presence of weak
  electron-electron scattering.  In a sufficiently long wire the
  scattering leads to full equilibration of the electron distribution
  function in the frame moving with the electric current.  At non-zero
  temperature this equilibrium distribution differs from the one supplied
  by the leads.  As a result the contact resistance increases, and the
  quantized conductance of the wire acquires a quadratic in temperature
  correction.  The magnitude of the correction is found by analysis of the
  conservation laws of the system and does not depend on the details of
  the interaction mechanism responsible for equilibration.

\end{abstract}
\maketitle

Experimental studies on the dc transport of short quantum wires have shown
the quantization of their conductance in units of $2 e^2/h$ \cite{expts}.
This phenomenon is well understood within a model of non-interacting
electrons \cite{nonint}, even though the interactions in the wire are
usually not weak, i.e., $e^2/\hbar v_F\gtrsim1$, where $v_F$ is the Fermi
velocity in the wire.  The absence of any effect of electron interactions
on the conductance is usually attributed to the fact that the quantum
wires are always connected to two-dimensional leads, where interactions
between electrons do not play a significant role.  Indeed, it has been
shown that within the so-called Luttinger-liquid theory, the interactions
inside the wire do not affect conductance \cite{maslov, ponomarenko,
  safi}.

A number of recent experiments revealed deviations from perfect
quantization in low-density wires \cite{thomas1, kane, thomas2,
  kristensen, thomas3, reilly, cronenwett, depicciotto, crook, rokhinson}.
These deviations often take the form of a shoulder-like feature, which
develops at finite temperature just below the first quantized plateau,
around $0.7 \times \left( 2 e^2 / h \right)$.  At the moment, there is no
consensus on the theoretical interpretation of this so-called ``$0.7$
structure.''  It is generally accepted, however, that electron-electron
interactions are involved in this feature, thus generating a lot of
interest in understanding the effect of interactions on the transport
properties of one-dimensional conductors.  Here we study one of the most
fundamental aspects of interactions, when they are so weak that their only
effect is to equilibrate inside the wire the electrons coming from the two
leads.

In the absence of interactions, the electrons propagate through the wire
ballistically.  Therefore the distribution functions of the right- and
left-moving electrons are controlled by the left and right leads,
respectively,
\begin{equation} \label{unperturbed}
f_p = \frac{\theta(p)}{e^{(\epsilon_p - \mu_{l})/T}+1}
     +\frac{\theta(-p)}{e^{(\epsilon_p - \mu_{r})/T}+1}.
\end{equation}
Here $\epsilon_p$ is the energy of an electron with momentum $p$,
$\theta(p)$ is the unit step function, and we assume that the left and
right leads have the same temperature $T$, but different chemical
potentials $\mu_l = \mu +eV$ and $\mu_r = \mu$.  It is important to note
that even weak processes of electron-electron scattering will modify the
distribution (\ref{unperturbed}).  Indeed, such processes will force some
left-moving electrons to change their direction of motion and become
right-movers.  Thus the basic assumption of Eq. (\ref{unperturbed}) that
all the right-movers originate from the left lead and are in equilibium
with it, will be violated.

The exact shape of the true steady state distribution of electrons can be
understood easily if the wire is very long and the electron system is
Galilean invariant, $\epsilon_p=p^2/2m$.  In this case it is convenient to
view the electron system in the reference frame moving with the drift
velocity $v_d=I/ne$, where $I$ is the electric current in the system and
$n$ is the electron density.  In this frame the electron system is at
rest, and must be described by the equilibrium Fermi-Dirac distribution
characterized by a single chemical potential.  In the stationary reference
frame this distribution takes the form
\begin{equation} \label{distrib}
f_p = \frac{1}
      {\exp\left(\frac{\epsilon_p - v_d p - \mu_{\rm eq}}{T_{\rm eq}}\right)+1},
\end{equation}
where $\mu_{\rm eq}$ and $T_{\rm eq}$ approach $\mu$ and $T$ at $V\to0$.
The distribution functions (\ref{unperturbed}) and (\ref{distrib})
coincide only in the limit of zero temperature. (In this case, to linear
order in the drift velocity, $\mu_{l,r}=\mu_{\rm eq}\pm v_d p_F$, where
$p_F$ is the Fermi momentum.)  At nonzero temperature, one can expect the
full equilibration of the electron system in the wire to significantly
affect its transport properties.

\begin{figure}[tb]
 \resizebox{.45\textwidth}{!}{\includegraphics{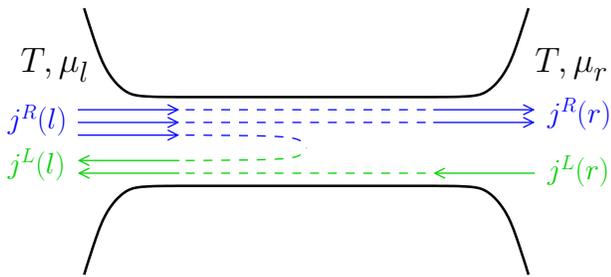}}
 \caption{\label{wire} Quantum wire in the regime of small applied bias,
   $\mu_l-\mu_r=eV$. The electric current is given by total currents of
   the right- and left-moving electrons, $I=e(j^R+j^L)$.  The
   equilibration processes convert some of the right-moving electrons into
   the left-moving ones, thereby reducing the conductance of the wire.}
\end{figure}

Recently, the equilibration of the electrons in the moving frame was shown
to have a strong effect on the Coulomb drag between two parallel quantum
wires \cite{pustilnik} and to give rise to a finite resistivity of long
inhomogeneous wires \cite{resistivity}.  On the other hand, because the
equilibration processes that relax the distribution (\ref{unperturbed}) to
the form (\ref{distrib}) involve converting right-moving electrons into
left-moving ones, it is natural to expect that the interactions will
affect conductance of the wires even in the absence of inhomogeneities,
Fig.~\ref{wire}.  Such an effect was recently discussed by Lunde,
Flensberg, and Glazman \cite{lunde} in the case of short wires, where the
effect of the equilibration processes is weak, and the distribution
function is still close to the unperturbed form (\ref{unperturbed}).  They
found a negative correction to the quantized conductance of the wire,
which grows linearly with its length $L$.

In this paper we explore the opposite limit of a long wire, $L\to\infty$,
in which the electron distribution function in the wire does assume the
limiting form (\ref{distrib}), and the correction to the conductance
saturates at a value independent of $L$.  The crossover between this
regime and that of short wires will be discussed elsewhere
\cite{unpublished}.

Throughout this paper we assume that the interactions between electrons
are very weak, and their only effect is to provide a mechanism of
relaxation of the distribution function to the form (\ref{distrib}).  The
exact nature of the scattering processes is unimportant, as long as these
processes conserve the number of electrons, the energy of the system, and
its momentum.  Below we obtain the conductance of the wire by detailed
analysis of these conservation laws.

The conservation of the number of particles means that in the steady state
regime the total particle current $j(x)$ is constant along the wire.  It
is convenient to present the total current as the sum $j=j^R+j^L$ of
currents of the right- and left-moving electrons,
\begin{equation} \label{jparticle}
j^{R,L}(x) = 2\int_{-\infty}^{\infty} \frac{dp}{h}\,\theta(\pm p) v_p f_p(x), 
\end{equation}
where the factor of 2 accounts for the spins, $v_p =p/m$ is the electron
velocity, positive sign in the step function corresponds to $j^R$, while
the negative one to $j^L$.

It is important to realize that the distribution function $f_p$ in
Eq.~(\ref{jparticle}) depends on the position in the wire.  Inside a long
wire, the relaxation processes ensure that $f_p$ has the universal form
(\ref{distrib}), but near the ends of the wire $f_p$ is affected by the
leads.  For example, at the left lead the distribution of the right-moving
electrons is controlled by the lead and takes the form of the first term
in Eq.~(\ref{unperturbed}).  This enables one to easily evaluate the
current $j^R(l)$ of the right-movers at the left lead.

Unlike the total current $j$, the current $j^R(x)$ is not uniform along
the wire, as the equilibration processes allow electrons to change
direction.  The rate $\dot{N}^R$ of the change of the number of
right-movers due to the electron-electron collisions is given by the
difference of the values of $j^R$ at the two ends of the wire, $\dot{N}^R
= j^R (r) - j^R (l)$.

Although the current $j^R(r)$ of the outgoing right-movers is not known,
it can be expressed in terms of the total current $j$ and the current
$j^L(r)$ of incoming left-movers, $j^R(r)=j-j^L(r)$.  It follows then that
the change in the number of right-movers per unit time $\dot{N}^R$ now
depends on the electric current $I=ej$ flowing through the wire, as well as
the sum of incoming particle currents from both leads
\begin{equation} \label{pflows}
j^R (l) + j^L (r) = \frac{I}{e} - \dot{N}^R .
\end{equation}
In analogy with $j^R(l)$, the current $j^L(r)$ is controlled by the right
lead and can be found by using the distribution function $f_p$ given by
the second term in Eq.~(\ref{unperturbed}).  Since both terms in the
left-hand side of Eq.~(\ref{pflows}) are determined by the distribution
functions in the non-interacting leads, the result of the routine
evaluation of the two currents is given by the Landauer formula
$j^R(l)+j^L(r) = 2eV/h$, up to corrections exponentially small in $\mu/T$.
Thus we find the following relation between the applied bias, electric
current, and $\dot{N}^R$,
\begin{equation} \label{nrdot}
\frac{2e^2 }{h} V = I - e\dot{N}^R .
\end{equation}
An equivalent relation was obtained earlier in Ref.~\cite{lunde} using the
Boltzmann equation formalism.  It formally expresses the idea that the
processes changing the number of right-movers in the wire will result in a
correction to the quantized conductance.

The conservation of energy in electron-electron collisions implies that
the total energy current $j_E(x)$ is uniform along the wire.  It is
instructive to express the energy current as the sum $j_E=j_E^R +j_E^L$ of
the contributions of the right- and left-moving particles,
\begin{equation} \label{ecurrent}
j^{R,L}_E(x) = 2\int_{-\infty}^{\infty} 
            \frac{dp}{h}\,\theta(\pm p) v_p\epsilon_p f_p(x).
\end{equation}

In the same fashion that we could relate the particle current $j$ to how
the number of right-moving electrons changes over time, one can find a
relation between the energy current $j_E$ flowing through the wire and the
rate of change $\dot{E}^R$ of the energy of right-movers due to the
electron collisions.  Indeed, the reasoning that led to Eq.~(\ref{pflows})
can be readily extended to the case of energy currents rather than
particle ones, leading to
\begin{equation} \label{eflows}
j_E^R (l) + j_E^L (r) = j_E - \dot{E}^R .
\end{equation}
The energy currents in the left-hand side of Eq.~(\ref{eflows}) are again
controlled by the leads and can be easily computed using the distribution
function (\ref{unperturbed}) of non-interacting electrons.  At low
temperatures $T \ll \mu$ one finds $j_E^R(l) + j_E^L(r) = (2eV/h)\mu$, up to
corrections small as $e^{-\mu/T}$.

Since the energy current $j_E$ does not depend on the position in the
wire, one can calculate it in a region away from the leads, where the
distribution function is given by Eq.~(\ref{distrib}).  The formal
calculation gives $j_E=\mu(1+\pi^2T^2_{\rm eq}/6\mu^2_{\rm eq})j$, where
we discarded terms of order $O\left((T_{\rm eq}/\mu_{\rm eq})^4\right)$
and higher.  This result can be compared with the above-mentioned
calculation for the unperturbed distribution (\ref{unperturbed}), which
can be summarized as $j_E=\mu j$.  The difference can be traced back to
the energy dependence of the term $v_dp$ in the Fermi function
(\ref{distrib}).  Expressing the particle current as $j=I/e$ we then
find
\begin{equation} \label{erdot}
\frac{2e^2}{h}V  =  \left[ 1 + \frac{\pi^2}{6} 
                   \left( \frac{T}{\mu}\right)^2 \right]I
                      -\frac{e}{\mu}\,\dot{E}^R.
\end{equation}
This result is obtained in the linear order in the applied bias, which
enabled us to substitute $T_{\rm eq}=T$ and $\mu_{\rm eq}=\mu$.

\begin{figure}[tb]
\begin{center}
\includegraphics[width=82mm]{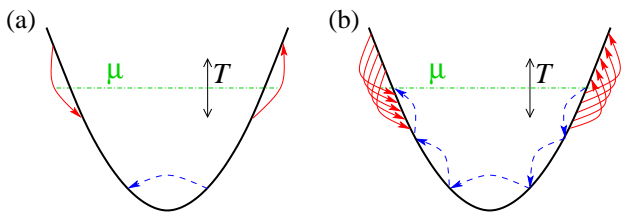} 
\end{center}
\vspace{-15pt} 
\caption{Illustration of the relaxation processes converting a
  right-moving electron into a left-moving one.  The parabolas represent
  the quadratic dispersion $\epsilon_p=p^2/2m$ of the electrons in the
  wire.  (a) A three-particle process studied in Ref.~\cite{lunde}.
  Momentum and energy conservation ensure that as a right-mover changes
  direction, another right-mover increases its energy and a left-mover
  decreases it.  (b) Transfer of an electron from the right to the left
  Fermi point is accompanied by multiple right-movers increasing their
  energies and multiple left-movers decreasing energies.}
\label{fig2}
\end{figure}

In the absence of the scattering processes changing the number of the
right- and left-moving electrons, not only $\dot N^R$, but also $\dot E^R$
would vanish.  Indeed, in this case the two branches of excitations would
represent two electron systems with no particle exchange allowed.  Then
the distribution function (\ref{unperturbed}) would describe the systems
of right- and left-movers in thermal equilibrium with each other, and net
heat transfer between them would vanish.  Thus  both $\dot{N}^R$ and
$\dot{E}^R$ arise as a consequence of the same relaxation mechanism inside
the wire, and we will now show that there is a simple relation between
these rates.

In the case of a short wire it was shown \cite{lunde} that the dominant
process changing the number of right-movers involves three electrons, with
a right-mover near the bottom of the band reducing its momentum and thus
the direction of motion, Fig.~\ref{fig2}(a).  The conservation of momentum
then requires that the other two electrons increase their momenta.
Finally, conservation of energy requires one of these two electrons to be
near the right Fermi point, and the other near the left one.  The typical
momentum change is controlled by the temperature, $|\delta p|\sim T/v_F$.
As a result of such scattering events the distribution function
(\ref{unperturbed}) shows only a small modification whereby the
exponentially small discontinuity near $p=0$ is smeared.


A much more significant change occurs in longer wires, where the
relaxation processes bring the distribution function to the form
(\ref{distrib}).  A comparison of the distribution functions
(\ref{unperturbed}) and (\ref{distrib}) shows that the main difference
between them is at the values of momentum $p$ near the Fermi points $\pm
p_F$.  Thus the dominant relaxation processes contributing to $\dot N^R$
take electrons with $p\approx p_F$ and move them to $p\approx -p_F$.  Such
processes are realized in many small steps of $|\delta p|\sim T/v_F$ and
are accompanied by multiple electrons scattering near the two Fermi
points, see Fig.~\ref{fig2}(b).  The total momentum transferred to these
electrons is $2p_F$.  Energy conservation requires that it is distributed
evenly between the right- and left-movers, so that the resulting energy
increase $\delta E^R=v_F p_F$ is compensated by the decrease $\delta
E^L=-v_F p_F$.

In the end, the energy balance for the right-moving electrons consists of
a loss of $\epsilon_F$, which was the energy of the electron changing
direction, and a gain of $v_F p_F=2 \epsilon_F$ due to the redistribution
of momentum.  As a result, for every right-moving electron that changes
direction, $\Delta N^R=-1$, the right-movers' energy increases by an
amount $\Delta E^R=\epsilon_F$.  We thus conclude that
\begin{equation} \label{statistics}
\dot{E}^R = - \mu \dot{N}^R,
\end{equation}
where we replaced $\epsilon_F$ with $\mu$, as the small difference
$\mu-\epsilon_F\sim T^2/\mu$ turns out to be irrelevant for our purposes.
It is important to note that the result (\ref{statistics}) is not
sensitive to the specific details of the electron relaxation mechanism.
Indeed, the two key ingredients of this derivation are the conservation
laws that control the redistribution of momentum $2p_F$ between the right-
and left-movers, and the quadratic dispersion that governs how the
energies of the two subsystems change as a result of that redistribution.

By analyzing the conservation laws we have so far been able to establish
three linear relations (\ref{nrdot}), (\ref{erdot}), and
(\ref{statistics}) between four quantities, $V$, $I$, $\dot N^R$, and
$\dot E^R$.  Assuming that the applied bias $V$ is known, we can now
express the remaining three quantities in terms of $V$.  Most importantly,
we find $I=GV$, with the conductance
\begin{equation} \label{conductance}
G = \frac{2e^2}{h} 
    \left[ 1 - \frac{\pi^2}{12} \left( \frac{T}{\mu} \right)^2\right],
\end{equation}
where we restricted ourselves to the leading order term in $(T/\mu)^2$.

The quadratic in temperature correction to the quantized conductance
$2e^2/h$ of the wire is our main result.  Unlike the correction to the
conductance of a short wire $\delta G\propto e^{-\mu/T} $\cite{lunde}, our
correction shows power-law dependence on the temperature.  Earlier papers
\cite{maslov, ponomarenko, safi} on the conductance of long quantum wires
did not find any correction to the conductance, as the Luttinger-liquid
theory used there does not account for the relaxation processes leading to
our result (\ref{conductance}).

Experimentally, small temperature-dependent corrections to quantized
conductance have been observed in quantum point contacts \cite{thomas1,
  thomas2, kristensen, cronenwett}.  The latter are essentially short
quantum wires, with only a few electrons in the one-dimensional part of
the device.  In order for our result (\ref{conductance}) to be fully
applicable the length of the system should be sufficient to ensure full
equilibration of the electron distribution function.  Comparison of our
correction $\delta G\sim (T/\mu)^2$ with the result \cite{lunde} for short
wires, $\delta G\propto Le^{-\mu/T}$ implies that our result is applicable
at $L\gg l_{\rm eq}\propto e^{\mu/T}$, in agreement with the more detailed
calculation \cite{unpublished}.  Although some experiments with longer
quantum wires have been reported \cite{kane,reilly}, a careful study of
the temperature dependent corrections to the conductance is not yet
available.

In summary, we have shown that in a long quantum wire, the full
equilibration of the electron distribution function leads to a finite
correction to the conductance, which at $T\ll \mu$ is quadratic in
temperature, Eq.~(\ref{conductance}).  Our derivation relied uniquely on
an analysis of the conservation laws for energy, momentum, and particle
number, without making specific assumptions regarding the process of
equilibration.

We are grateful to A. V. Andreev and L. I. Glazman for helpful
discussions.  This work was supported by the U.S. Department of Energy,
Office of Science, under Contract No. DE-AC02-06CH11357.

\end{document}